# Scatterplot Selection Applying a Graph Coloring Problem

Takayuki Itoh[1)]    Asuka Nakabayashi[1)]    Mariko Hagita[1)]
1) Ochanomizu University

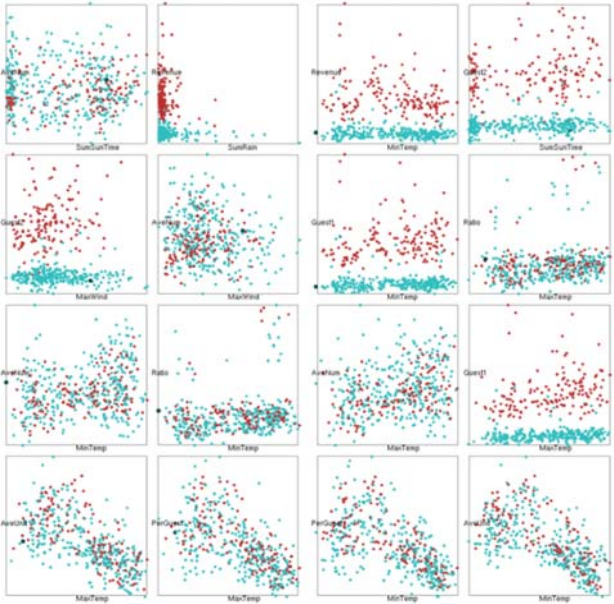

Figure 1. Visualization example by our technique. Several scatterplots show strong correlations between dimension pairs, several ones clearly show clusters or outliers, while several others show how two labels drawn in red and blue are separated. Our technique selects a variety of scatterplots to show various characteristics of the input multidimensional dataset in a single display space.

## ABSTRACT

Scatterplot selection is an effective approach to represent essential portions of multidimensional data in a limited display space. Various metrics for evaluation of scatterplots such as scagnostics have been presented and applied to scatterplot selection. This paper presents a new scatterplot selection technique that applies multiple metrics. The technique firstly calculates scores of scatterplots with multiple metrics and then constructs a graph by connecting similar scatterplots. The technique applies a graph coloring problem so that different colors are assigned to similar scatterplots. We can extract a set of various scatterplots by selecting them that the specific same color is assigned. This paper introduces visualization examples with a retail dataset containing multidimensional climate and sales values.

## 1 INTRODUCTION

Multidimensional data visualization has been one of the most active research issues in the visualization community. There have been various techniques on multidimensional data visualization including geometric techniques such as scatterplot matrix (SPM) and parallel coordinate plots (PCP), iconic techniques, and pixel-based techniques. Dimension selection [5,19,21] is one of the most important issues for the visualization of high-dimensional data. It is not reasonable to represent every dimension in a limited display space; therefore, it is important to remove noisy or meaningless dimensions and to focus on the visualization of informative dimensions.

Many recent studies on multidimensional data visualization presented a variety of metrics that denote the informativeness of scatterplots. Scagnostics [18] is one of the typical metrics for the scatterplots and has been applied to scatterplot selection problems [10,17,21]. Meanwhile, it is not always appropriate to apply a single metric for scatterplot selection to represent the overall characteristics of the multidimensional data. For example, interesting correlations are observed from some pairs of dimensions while interesting clusters are observed from some other pairs of dimensions. In this case, we may want to display scatterplots that have various characteristics in a single display space.

This paper presents a new and fast technique for scatterplot selection with multiple metrics. This technique firstly generates scatterplots with arbitrary pairs of dimensions. Then, multiple scores based on multiple metrics are calculated for each scatterplot and a vector is formed from the scores. The technique constructs a graph by connecting pairs of scatterplots if the similarity between their vectors is larger than a user-defined threshold. It then assigns colors to the vertices corresponding to the scatterplots while complying with a rule that different colors are assigned to a pair of vertices connected by an edge. In other words, the same color is assigned to a set of significantly different scatterplots. The technique selectively displays a constant number of scatterplots that have the same color. As shown in Figure 1, the technique realizes the selection of a variety of scatterplots that show various characteristics of the input dataset.

This paper introduces a case study with a consumer business dataset including climate and revenue values.

## 2 RELATED WORK

### 2.1 Dimension Selection for Multi-dimensional Data Visualization

Dimension selection techniques have been widely applied to multi-dimensional data visualization so that an important subset of dimensions can be effectively represented. Claessen et al. [2] visualized high-dimensional datasets by representing a set of low-dimensional subspaces as a combination of PCPs and scatterplots. Suematsu et al. [15] and Zheng et al. [22] also converted high-dimensional datasets into low-dimensional subsets and visualized these subsets using multiple PCPs or scatterplots respectively. These techniques did not provide rich interaction mechanisms to freely select the numbers of dimensions.

Several recent studies have demonstrated interaction mechanisms to freely visualize interesting low-dimensional subspaces. Lee et al. [6] and Liu et al. [7] applied dimension reduction schemes to interactively select subsets of the high-dimensional data. Nohno et al. [11] presented a technique to interactively contract highly-correlated dimensions to adjust the number of axes displayed in PCPs. Itoh et al. [5], Watanabe et al. [17] and Nakabayashi et al. [10] presented a series of techniques that easily control the number of dimensions displayed in the PCPs or number of dimension pairs represented by scatterplots.

It is also important to understand relationships among dimensions while extracting low-dimensional subspaces. Dimension spaces have been visualized by applying scatterplots or graphs by several recent studies [10,17,21]. This is also an effective approach to interactively select meaningful sets of dimensions.

Despite a lot of studies on multidimensional data visualization have applied dimension selection techniques, there have been few studies to automatically select a variety of a limited number of informative scatterplots. We address this problem and present a new technique in this paper.

## 2.2 Evaluation of Scatterplots

Numeric evaluation of the informativeness of scatterplots has been an active research topic. Scagnostics is a famous concept to quantitatively evaluate the informativeness of scatterplots. Wilkinson et al. [18] proposed nine features of scagnostics based on the appearance of the scatterplots. Wang et al. [16] proposed the improved scagnostics by considering human perception to several metrics including "Outlying" and "Clumpy." There have been more several more studies that focus on specific metrics of scatterplots, including correlation [4,13] and class separation [1,12,14]. There have been several visualization studies on overview and exploration of a large number of scatterplots. Dang et al. [3] presented an exploration mechanism for finding similar scatterplots and filtering scagnostics. Matute et al. [8] presented another approach to representing the distribution of characteristics of scatterplots. The goal of our study is somewhat similar to the above studies since we also focus on presenting a variety of scatterplots; however, our focus is different from these studies since we aim to select the fixed number of a variety of scatterplots.

## 3 SCATTERPLOT SELECTION APPLYING A GRAPH COLORING PROBLEM

This section presents a processing flow of the presented scatterplot selection technique. We suppose that scores of scatterplots are calculated with multiple metrics and stored as vector values. Figure 2 illustrates the concept of scatterplot selection. Scatterplots are depicted as vectors in the metrics space. The requirements for scatterplot selection in this study are summarized as follows:

[R1:] Select distant and long vectors to select a variety of informative scatterplots.

[R2:] Avoid selecting multiple close vectors to avoid selecting similar scatterplots.

[R3:] Avoid selecting short vectors to avoid selecting less informative scatterplots.

We present a graph coloring problem to satisfy the above requirements and display a variety of informative scatterplots.

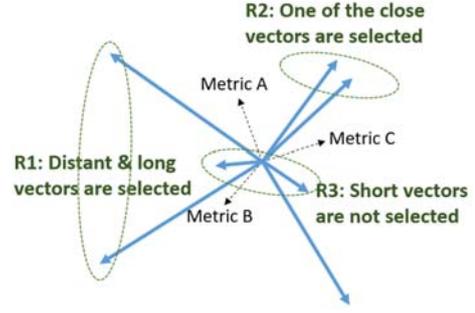

Figure 2. Concept of scatterplot selection in the metrics space. Blue arrows illustrate the vectors of metrics. Our technique selects a variety of scatterplots while satisfying **R1**, **R2** and **R3**.

### 3.1 Data Structure

This paper formalizes the problem as follows. An input multidimensional dataset $A$ has $n$ individuals as
$$A = \{a_1, a_2, \ldots, a_n\}.$$
The i-th individual $a_i$ has the $m$-dimensional values as
$$a_i = (a_{i1}, a_{i2}, \ldots, a_{im}).$$
A set of scatterplots formed from arbitrary pairs of dimensions is described as
$$S = \{s_1, s_2, \ldots, s_N\},$$
where $N$ is the total number of scatterplots. Each scatterplot has a set of scores calculated based on predefined metrics, described as
$$s_i = (s_{i1}, s_{i2}, \ldots, a_{iM}),$$
where $M$ is the number of metrics. The cosine similarity between the $i$-th and the $j$-th scatterplots is described as
$$d_{ij} = (s_i \cdot s_j)/(|s_i||s_j|).$$

### 3.2 Metrics for Scatterplot Selection

Our technique calculates multiple scores for each scatterplot based on multiple metrics. Our current implementation supports the following metrics.

#### 3.2.1 Correlation

Correlation is one of the most common metrics to determine the relationship between a pair of dimensions. Our current implementation just calculates the score of the $k$-th scatterplot as follows:
$$s_{k1} = |S_{pear}(i,j)^2|$$
where $S_{pear}(i,j)$ is the Spearman's rank correlation between the $i$-th and the $j$-th dimensions. A dimension pair gets a higher score if they have a strong positive/negative correlation.

Newer approaches on correlation [4,13] can be also applied.

#### 3.2.2 Thinness

It means easier to adopt a mathematical model to a set of individuals if they form thin regions in a scatterplot. We measure the thinness of the region where the individuals place in the scatterplot as Wilkinson et al. [18] did. Our implementation generates a Delaunay triangular mesh $T$ connecting the individuals in a scatterplot and then removes all triangles which have at least one edge that is longer than a pre-defined threshold. Then. we calculate the score as follows:
$$s_{k2} = 1 - \sqrt{4\pi A_{rea}(T)}/P_{erimeter}(T)$$
where $A_{rea}(T)$ is the total area of $T$, and $P_{erimeter}(T)$ is the total length of the boundary of $T$.

### 3.2.3 Clumping

It is remarkable if the individuals in a scatterplot are well-separated into several clusters. Our current implementation simply applies the metric "Clumpy" presented by Wilkinson et al. [18] defined as follows:

$$s_{k3} = 1 - length(e_{maxr})/length(e_{mind})$$

Here, our implementation generates a Delaunay triangular mesh as described in the previous section, and deletes the edges that are longer than $e_{mind}$. Meanwhile, $e_{maxr}$ is the longest remaining edge.

Newer approaches on clumping [16] can be also applied.

### 3.2.4 Separateness

Suppose that one of the labels is assigned to each of the individuals. It is remarkable if the individuals that have a particular same label are well-separated in a scatterplot. We measure the separateness of a particular label by calculating the entropy of the labels. In particular, we compute the entropy of the labels in the scatterplot generated with the *i*-th and the *j*-th dimensions as follows:

$$H(i,j) = -\frac{1}{n}\sum_{k=1}^{n}\sum_{c=1}^{C} p(y_k = c|(a_{ki}, a_{kj})) \log p(y_k = c|(a_{ki}, a_{kj}))$$

where $y_k$ is the label of the *k*-th individual, $(a_{ki}, a_{kj})$ is the position in the scatterplot of the *k*-th individual, and $C$ is the number of labels.

Our implementation divides the scatterplot into $L$ subareas and calculate the entropy at the *l*-th subarea $H(i,j)_l$ by the above equation, and finally calculates the score of the *k*-th scatterplot as follows:

$$s_{k4} = (H_{max} - \sum H(i,j)_l)/H_{max}$$

where $H_{max}$ is the maximum value of $\sum H(i,j)_l$.

Other approaches [1,12] can be also applied to determine the class separateness.

## 3.3 Graph Coloring Problem

This technique applies a graph coloring problem to select a variety of scatterplots that have different characteristics. This idea is originally presented for the selection of a variety of photos from a large-scale collection [9]. Suppose a graph $G=\{S,E\}$, where $S$ is a set of vertices corresponding to the scatterplots, and $E$ is a set of edges connecting pairs of scatterplots. The technique constructs the graph by generating edges between the *i*-th and the *j*-th scatterplots if their similarity $d_{ij}$ is larger than the pre-defined threshold $d_{thres}$.

Then, the technique assigns colors to the scatterplots while complying a rule that different colors are assigned to a pair of vertices connected by an edge. In other words, the same color is assigned to a set of significantly different scatterplots. Figure 3 illustrates the process. The process firstly selects the scatterplot that have the largest $s_k$, and assigns the color identification $c_k$=0. Then, adjacent vertices connected by edges are traversed in the breadth-first order. While visiting the *k*-th vertex, the process specifies the minimum color identification that is assigned to none of the adjacent vertices connected with the *k*-th vertex, and assigns it to the *k*-th vertex. For example, if color identifications 0, 1, and 3 have been assigned to the vertices adjacent to $c_k$, the process specifies $c_k$ as 2. The breadth-first search is repeated until color identifications are assigned to all the vertices.

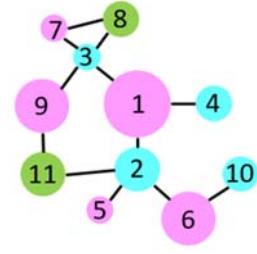

Figure 3. Graph coloring. The process assigns different colors to the vertex pairs connected by edges. The numbers in this figure denote the order of the breadth-first search.

Finally, we select a predefined number of scatterplots to be displayed. The technique extracts a set of scatterplots in which the same color is assigned. We calculate the sums of the length of the vectors $s_k$, for each color and select the color that brings the largest sum. The extracted set of scatterplots does not include any similar pairs because similar pairs of scatterplots are connected and therefore have different colors. In other words, it satisfies **R1** and **R2** because the extracted set consists of a variety of differently looking scatterplots. If the number of extracted scatterplots is larger than the predefined number, the technique selects the scatterplots in the descending order of max($s_{k1}$, $s_{k2}$, $s_{k3}$, $s_{k4}$), the maximum value of the four scores, to satisfy **R1** and **R3**.

The processing flow is as follows.

1. Initialize the vertices $S$. Calculate the interestingness of the *k*-th scatterplot as $s_k$,.
2. Construct the graph. Generate an edge between the *i*-th and the *j*-th scatterplots if $d_{ij}$ is larger than the pre-defined threshold.
3. Select the scatterplot that has the largest interestingness as the starting vertex.
4. Traverse the connected vertices by the breadth-first search. Assign color identifications to the traversed vertices. Repeat this traverse until the color identifications are assigned to all the vertices.
5. Collect the vertices that have the same color identification. Select the predefined number of vertices in the descending order of max($s_{k1}$, $s_{k2}$, $s_{k3}$, $s_{k4}$).

## 4 EXAMPLE

This paper introduces an example of visualization by the presented technique applying a retail transaction and climate dataset. Table 1 shows the explanatory variables (climate values) assigned to the horizontal axis while Table 2 shows the objective functions (retail transaction values) assigned to the vertical axis in this dataset. The dataset contained the records of 457 days from May 1, 2016, to July 31, 2017, corresponding to 457 data points in the scatterplots. We generated 35 scatterplots consisting of five horizontal axes and seven vertical axes. Remark that this dataset is perturbed by adding random small real values to each column of the original dataset. The data points are drawn in red or blue: red denotes holidays while blue denotes weekdays.

Figure 1 shows an example of scatterplot selection by our technique. Here, several scatterplots show correlations between dimension pairs, several ones show clusters or outliers, while several others show how two labels drawn in red and blue are separated. This figure demonstrates that our technique successfully selects a variety of scatterplots to show various characteristics of the dataset.

Table 1: The explanatory variables (climate values).

| MinTemp | Minimum temperature |
|---|---|
| MaxTemp | Maximum temperature |
| SumRain | Precipitation |
| SumSunTime | Sunshine duration |
| MaxWind | Maximum wind speed |

Table 2: The objective functions (retail transaction values).

| Revenue | Revenue |
|---|---|
| Guest1 | Number of customer |
| Guest2 | Number of visitor |
| Ratio | Conversion rate |
| PerGuest | Average revenue per customer |
| AveUnit | Average price of purchased items |
| AveNum | Average number of purchased items |

Figures 4, 5 and 6 show top four scatterplots that archived the highest scores on correlation, entropy, and clumpy. Here, the horizontal axes of scatterplots are MinTemp or MaxTemp while the vertical axes are PerGuest or AveUnit in Figure 4. It denotes that the average revenue or price well correlates with the temperature. Meanwhile, the vertex axes of scatterplots in Figure 5 are Revenue, Guest1, or Guest2. It denotes that revenue and the number of guests drastically different between holidays and weekdays. Figure 6 suggests a set of dimension pairs that bring better views to discover outliers and clusters. The scatterplot selection result shown in Figure 1 is well-balanced because it represents various characteristics of the input dataset by selecting various scores of scatterplots. Meanwhile, Figure 7 shows examples of scatterplots that have no higher scores with all the metrics. Actually, these scatterplots do not look characteristic or informative. The presented technique does not aggressively select such types of scatterplots.

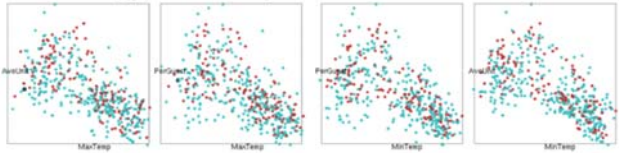

Figure 4. Scatterplots that archived the highest scores on correlation.

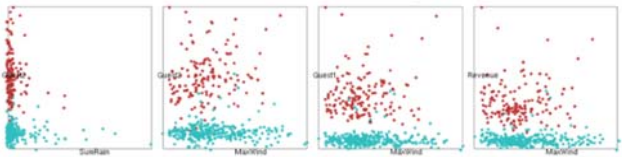

Figure 5. Scatterplots that archived the highest scores on entropy.

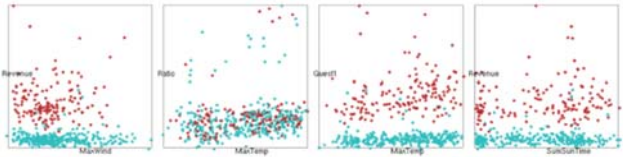

Figure 6. Scatterplots that archived the highest scores on clumpy.

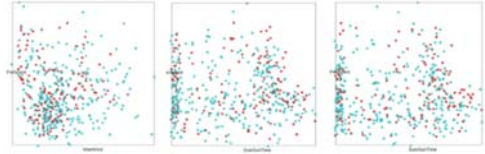

Figure 7. Examples of scatterplots that have no higher scores with all metrics.

Figure 8 shows the normalized scores of four metrics of sixteen scatterplots shown in Figure 1. It demonstrates the selection of various scores of scatterplots.

The result of scatterplot selection strongly depends on the choice of $d_{thres}$. The smaller $d_{thres}$ brings a larger number of edges and consequently a larger number of scatterplots groups corresponding to the number of colors in Figure 3. Table 3 shows the numbers of edges and colors in our experiments. Here, the selection of very similar scatterplots would be avoided by making a larger number of groups, but at the same time, informativeness of the selected scatterplots may be decreased. Figure 9 shows the maximum scores $\max(s_{k1}, s_{k2}, s_{k3}, s_{k4})$ of the selected scatterplots while adjusting the $d_{thres}$ values. This result suggests that we need to carefully adjust this threshold to select a variety of informative scatterplots.

Table 3: The numbers of edges and colors.

| $d_{thres}$ | 0.9995 | 0.9975 | 0.995 | 0.9925 | 0.99 |
|---|---|---|---|---|---|
| Edges | 5 | 19 | 39 | 60 | 83 |
| Colors | 2 | 3 | 5 | 6 | 7 |

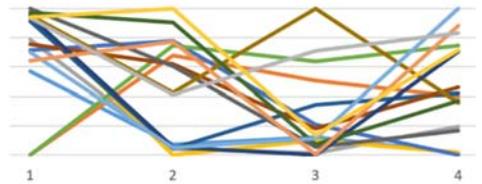

Figure 8. Scores of four metrics of sixteen scatterplots shown in Figure 1.

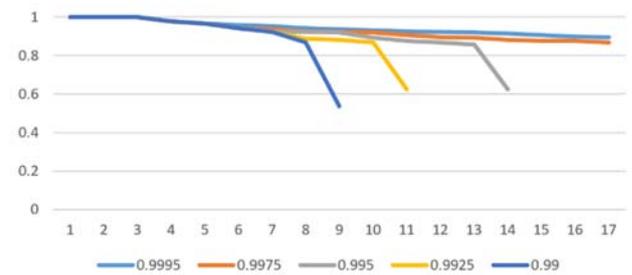

Figure 9. Ranks of maximum scores while adjusting $d_{thres}$.

## 5 CONCLUSION AND FUTURE WORK

This paper presented a new scatterplot selection technique applying a graph coloring problem. The technique calculates scores based on several independent metrics for each scatterplot. Then, the technique constructs a graph by connecting vertex pairs corresponding to scatterplot pairs if these scores are similar. The graph coloring problem is applied to the graph, and scatterplots that the user-specified color is assigned are extracted. The paper introduced examples of the scatterplot selection applying a retail transaction and climate dataset.

Our future issues include the following. Firstly, we would like to add and modify the metrics. There have been various improved metrics for scagnostics as mentioned in Section 2.2. We would like to apply them and explore the best combination of the metrics for this study. Then, we would like to test the scalability of the presented technique. Especially, we suppose it is necessary to test the datasets that have a large number of dimensions, and therefore a large number of scatterplots can be generated. Also, it is necessary to test the datasets with a large number of individuals. After the above improvements and tests, we would like to have case studies with various real-world datasets and conduct user evaluations.